\documentclass[prd,preprint,preprintnumbers,nofootinbib,eqsecnum,superscriptaddress]{revtex4}

 \usepackage[dvips,final]{graphicx}
  \usepackage{amssymb}
   \usepackage{amsmath}
    \usepackage{amsfonts}
     \usepackage{epsfig}
      \usepackage{bm}

\usepackage{mathpazo}

\usepackage[section]{placeins}

\usepackage{multirow}
\usepackage{ctable}
\usepackage{booktabs}
\usepackage{array}
\usepackage{tabularx}
\usepackage{xcolor}
\usepackage{pstricks}
\definecolor{fred}{rgb}{0.90053, 0.00369, 0.00159}  

\newcommand{\GeV}{\textrm{GeV}}

\newcommand{\bq}{\mbox{\boldmath $q$}}

\newcommand{\bk}{\mbox{\boldmath $k$}}

\begin{document}

\author{Anna Cisek}
\email{acisek@ur.edu.pl}
\affiliation{College of Natural Sciences, Institute of Physics,
University of Rzesz\'ow, ul. Pigonia 1, PL-35-959 Rzesz\'ow, Poland}

\author{Antoni Szczurek\footnote{also at University of Rzesz\'ow, 
PL-35-959 Rzesz\'ow, Poland}}
\email{antoni.szczurek@ifj.edu.pl} 
\affiliation{Institute of Nuclear
Physics, Polish Academy of Sciences, ul. Radzikowskiego 152, PL-31-342 Krak{\'o}w, Poland}

\title{Two-gluon production of $\phi$ and $\eta'$ mesons in
  proton-proton collisions at high energies}

\begin{abstract}
We discuss gluon-gluon mechanisms for production of mesons with hidden
strangeness, such as $\eta'$ and $\phi$ meson, in proton-proton
collisions at large energies.
The $g^* g^* \to \eta'$ and $g^* g^* \to \phi g$ mechanisms
are considered only and the corresponding cross sections are
calculated in the $k_t$-factorization approach. 
The $F_{\gamma^* \gamma^* \to \eta'}$ and $F_{g^* g^* \to \eta'}$ form
factors are calculated from quark-antiquark
$\eta'$ light-cone wave function including quark/antiquark transverse
momenta. The result for two-photon transition form factor demonstrates 
that higher twists may survive even to large photon virtualities.
The result is compared with the result of a recent leading-twist
NLO analysis which uses phenomenological distribution amplitudes
fitted to exclusive production of 
$\eta'$ in $e^+ e^- \to e^+ e^- \eta'$ reaction.
We calculate transverse momentum distributions of both $\eta'$ and
$\phi$ mesons in proton-proton collisions for RHIC and LHC energies. 
The results are compared to experimental data whenever available.
The results of the Lund string model are shown for comparison 
for $\eta'$ production at LHC energies.
It seems that the two-gluon fusion is not the dominant mechanism
for both $\phi$ and $\eta'$ production, although the situation for
$\eta'$, especially at larger energies, is less clear due to lack
of experimental data.
\end{abstract}

\maketitle

\section{Introduction}

The mechanism of the particle production in proton-proton
collisions was studied continously for five decades.
Pions and kaons are belived to be produced in the fragmentation
process. The Lund string model is state of art in this context.
On the other side quarkonium production at high energies is studied 
considering two-gluon process. Depending on $C$-parity we have
either $g^* g^* \to {\cal Q}$ ($C$ = +1) or $g^* g^* \to {\cal Q} g$
($C$ = -1) processes when limiting to color singlet mechanisms.
Color octet processes are not under full theoretical control.
Recently our group showed that the production of $\eta_c$ quarkonium
can be understood asumming simple color-singlet $g^* g^* \to \eta_c$
fusion \cite{BPSS2020}. The situation with resonances, such as
$\rho^0$, $f_0(980)$ or $f_2(1270)$ is still different.
We have shown that at large isoscalar meson transverse momenta
the gluon-gluon fusion may be an important mechanism
\cite{LMS2020,LS2020}. At low transverse momenta one has to
included also coalescence mechanism \cite{LS2020}.
For fully heavy tetraquark production the double-parton mechanism may be
required \cite{MSS2021}.

The situation with hidden strangness mesons was not discussed
carefully in the literature. In PYTHIA \cite{Pythia} such mesons 
are produced within the Lund-string model.
Recently there was some works on modification of strange hadron
production. Effects of the rope hadronization on strangeness enhancement
in $p p$ collisions at the LHC were discussed e.g. in \cite{BGLT2015}.

Here we wish to explore whether the two-gluon fusion
may be also important production mechanism.
It was suggested that the hadronic production of $\eta'$ meson
could be used to extract two-gluon transition form factor 
\cite{MY2000,AP2002}.
The formalism of $\eta'$ meson production in proton-proton collisions 
was discusssed already some time ago but no explicit calculation was
performed so far.
In addition, it could not be compared to
the data as the latter was not available at that time (2007).
In the meantime the PHENIX collaboration measured the $\eta'$
production at $\sqrt{s}$ = 200 GeV but no comparison was
made with theoretical results according to our knowledge.
Here we wish to study the situation more carefully and 
make comparison to realistic calculation of the gluon-gluon fusion.

Exclusive $p p \to p p \eta'$ production was studied in \cite{SPT2007}
within KMR perturbative approach and the measured cross section could
not be explained at the relatively low $\sqrt{s}$ = 29.1 GeV energy
of the WA102 collaboration experiment at CERN SPS.
On the other hand soft pomeron/reggeon exchanges can be fitted to
describe the experimental data \cite{LNS2014}.

The presence of the two-gluon component in $\eta'$ may be relevant
in the context of its production in the hadronic reaction.
The gluon content in a meson may occur e.g. via mixing with glueballs.
Mixing of scalar glueball with the scalar-isoscalar ``quarkonia'' was
discussed e.g. in \cite{GGF2005}.
According to our knowledge there was not such a discussion for $\eta'$.
However, according to lattice QCD pseudoscalar glueball has a mass
of about 2.6 GeV \cite{Chen2006,SFAPRX2020}, so the mixing 
should be rather small. Lower mass pseudoscalar glueball was discussed
in \cite{GLT2009}, which mixes however rather with radial excitations.
Review on experimental searches for the light pseudoscalar glueball
can be found in \cite{MCU2006}.
On the other hand axial anomaly may ``cause'' the presence of gluons
in the $\eta'$ wave function \cite{BM2019}.
No big gluonic content either in $\eta$ or $\eta'$ was found from
the $V \to P \gamma$ and $P \to V \gamma$ radiative decays in
\cite{EN2007}. The KLOE-2 collaboration found $Z_g^2 \approx$ 0.11
probability of the two-gluon component \cite{KLOE2}.

For comparison, the hadronization process for $\eta'$ production
was not discussed carefully in the literature.
Some discussion was presented e.g. in \cite{AGS1994} but in the context
of $e^+ e^-$ collisions.

In this paper we shall discuss possible consequences of gluonic component
in $\eta'$ for its production in proton-proton collisions.
We shall compare our results both with experimental data as well as
with results of the Lund string model.

\section{Sketch of the formalism}

\begin{figure}
\begin{center}
\includegraphics[width=6.5cm]{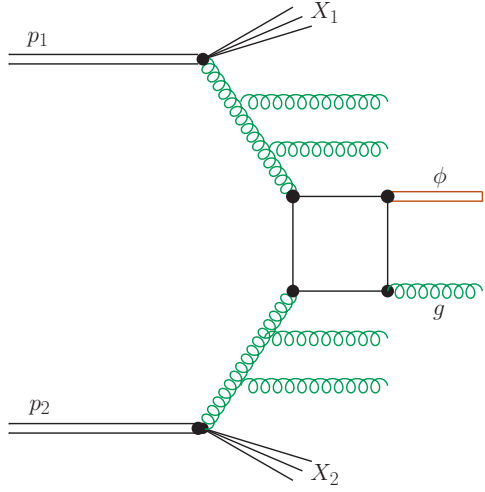}
\end{center}
\caption{The leading-order diagram for direct $\phi$ 
meson production in the $k_t$-factorization approach.}
\label{fig:gg_phig}
\end{figure}

The main color-singlet mechanism of $\phi$ meson production is illustrated
in Fig.\ref{fig:gg_phig}. In this case $\phi$ is produced
in association with an extra ``hard'' gluon due to C-parity conservation.

We calculate the dominant color-singlet $g g \to \phi g$
contribution taking into account transverse momenta of initial gluons.
In the $k_t$-factorization the NLO differential cross section can 
be written as:
\begin{eqnarray}
\frac{d \sigma(p p \to \phi g X)}{d y_{J/\psi} d y_g d^2 p_{\phi,t} d^2 p_{g,t}}
&& = 
\frac{1}{16 \pi^2 {\hat s}^2} \int \frac{d^2 q_{1t}}{\pi} \frac{d^2 q_{2t}}{\pi} 
\overline{|{\cal M}_{g^{*} g^{*} \rightarrow \phi g}^{off-shell}|^2} 
\nonumber \\
&& \times \;\; 
\delta^2 \left( \vec{q}_{1t} + \vec{q}_{2t} - \vec{p}_{H,t} - \vec{p}_{g,t} \right)
{\cal F}_g(x_1,q_{1t}^2,\mu^2) {\cal F}_g(x_2,q_{2t}^2,\mu^2) \; ,
\label{kt_fact_gg_jpsig}
\end{eqnarray}
where ${\cal F}_g$ are unintegrated (or transverse-momentum-dependent) 
gluon distributions.
The matrix elements were calculated as done e.g. for $J/\psi g$ production 
in \cite{CS2018}.
The corresponding matrix element squared for the $g g \to \phi g$ is
\begin{equation}
|{\cal M}_{gg \to \phi g}|^2 \propto \alpha_s^3 |R(0)|^2 \; .
\label{matrix_element} 
\end{equation}
Running coupling contants are used in the calculation. 
Different combination of renormalization scales were tried. 
Finally we decided to use:
\begin{equation}
\alpha_s^3 \to \alpha_s(\mu_1^2) \alpha_s(\mu_2^2) \alpha_s(\mu_3^2) \; ,
\end{equation}
where $\mu_1^2 = \max(q_{1t}^2,m_t^2)$,
      $\mu_2^2 = \max(q_{2t}^2,m_t^2)$ and
      $\mu_3^2 = m_t^2$,
where here $m_t$ is the $\phi$ transverse mass.
The factorization scale in the calculation was taken as
$\mu_F^2 = (m_t^2 + p_{t,g}^2)/2$.

The radial wave function at zero can be estimated from the decay
of $\phi \to l^+ l^-$ as is usually done for 
$J/\psi(c \bar c)$, see e.g. \cite{Mangoni}
\begin{equation}
\Gamma(\phi \to l^+ l^-) = 16 \pi \frac{\alpha Q_s^2}{M_{\phi}^2}
|\Psi_{\phi}(0)|^2 \left(1 - \frac{16}{3} \frac{\alpha_s}{\pi} \right)
\; ,
\label{Gamma_phi_from_Psi(0)}
\end{equation}
where $Q_s$ is fractional charge of the $s$ quark.
Then
\begin{equation}
|\Psi_{\phi}(0)|^2 = \frac{\Gamma(\phi \to l^+ l^-)}{16 \pi \alpha_{em} Q_s^2}
\frac{M_{\phi}^2}{1 - 16 \alpha_s/(3 \pi)} \; .
\label{Psi(0)}
\end{equation}
In the evalution we use $\alpha_s$ = 0.3.
Using branching fraction from PDG \cite{PDG2020} we get $|\Psi(0)|$.
By convention $|R(0)|^2 = 4 \pi |\Psi(0)|^2$.
Assuming $\alpha_s$ = 0.3 we get $|R(0)|^2$ = 0.11 GeV$^3$.
We shall use this value to estimate the cross section for production
of $\phi$ meson. For comparison for $J/\psi$ (real quarkonium) one gets
$|R(0)|^2 \approx$ 0.8 GeV$^3$.

\begin{figure}
\begin{center}
\includegraphics[width=6.5cm]{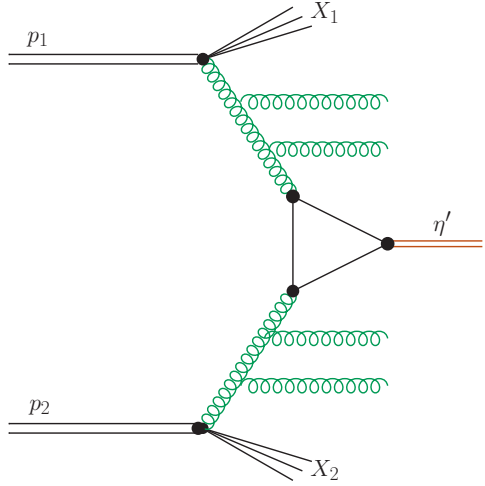}
\end{center}
\caption{The leading-order diagram for $\eta'$ meson production
in the $k_t$-factorization approach.}
\label{fig:gg_etap}
\end{figure}

Similarly we perform calculation for S-wave $\eta'$ meson production.
Here the lowest-order subprocess $g g \to \eta'$ is allowed by
positive $C$-parity of $\eta'$ mesons.
In the $k_t$-factorization approach the leading-order cross section 
for the $\eta'$ meson production can be written as:
%
%

%
\begin{eqnarray}
\sigma_{pp \to \eta'} = \int d y d^2 p_t d^2 q_t \frac{1}{s x_1 x_2}
\frac{1}{m_{t,\eta'}^2}
\overline{|{\cal M}_{g^*g^* \to \eta'}|^2} 
{\cal F}_g(x_1,q_{1t}^2,\mu_F^2) {\cal F}_g(x_2,q_{2t}^2,\mu_F^2) / 4
\; ,
\label{useful_formula}
\end{eqnarray}
that can be also used to calculate rapidity and transverse
momentum distribution of the $\eta'$ mesons.
Above  ${\cal F}_g$ are unintegrated (or transverse-momentum-dependent) 
gluon distributions and $\sigma_{g g \to \eta'}$ is 
$g g \to \eta'$ (off-shell) cross section.
In the last equation:
$\vec{p}_t = \vec{q}_{1t} + \vec{q}_{2t}$ is transverse momentum 
of the $\eta'$ meson
and $\vec{q}_t = \vec{q}_{1t} - \vec{q}_{2t}$ is auxiliary variable 
which is used in the integration. Furthermore:
$m_{t,{\eta'}}$ is the so-called $\eta'$ transverse mass and
$x_1 = \frac{m_{t,\eta'}}{\sqrt{s}} \exp( y)$,  
$x_2 = \frac{m_{t,\eta'}}{\sqrt{s}} \exp(-y)$.
The factor $\frac{1}{4}$ is the jacobian of transformation from
$(\vec{q}_{1t}, \vec{q}_{2t})$ to $(\vec{p}_t, \vec{q}_{t})$ variables.
The situation is illustrated diagrammatically in Fig.~\ref{fig:gg_etap}.

As for $\phi$ production the  running coupling contants are used. 
Different combination of scales are tried. The best choice is:
\begin{equation}
\alpha_s^2 \to \alpha_s(\mu_1^2) \alpha_s(\mu_2^2) \; ,
\end{equation}
where $\mu_1^2 = \max(q_{1t}^2,m_t^2)$ and
      $\mu_2^2 = \max(q_{2t}^2,m_t^2)$.
Above $m_t$ is transverse mass of the $\eta'$ meson.
The factorization scale(s) for the $\eta'$ meson production are 
fixed traditionally as $\mu_F^2 = m_t^2$.

The $g^* g^* \to \eta'$ coupling has relatively simple one-term form:
\begin{equation}
T_{\mu \nu}(q_1,q_2) = F_{g^* g^* \eta'}(q_1,q_2) 
\epsilon_{\mu \nu \alpha \beta} q_1^{\alpha} q_2^{\beta}  \; ,
\label{gg_etap_coupling}
\end{equation}
where $F_{g^* g^* \to \eta'}(q_1,q_2)$ object is known as the two-gluon
transition form factor.
The matrix element to be used in the $k_t$-factorization is then:
\begin{equation}
{\cal M}^{a b} = \frac{q_{1,\perp}^{\mu} q_{2,\perp}^{\nu}}
                      {|{\bf q}_1| |{\bf q}_2| } T_{\mu \nu} \; .
\end{equation}
In contrast to the convention for two-photon transition form factor
the strong coupling constants are usually absorbed into the two-gluon 
form factor definition.

The matrix element squared for the $g g \to \eta'$ subprocess is
\begin{equation}
|{\cal M}_{gg \to \eta'}|^2 \propto 
F_{g^* g^* \to \eta'}^2(q_{1t}^2.q_{2t}^2)
\propto \alpha_s^2 F_{\gamma^* \gamma^* \to \eta'}^2(q_{1t}^2.q_{2t}^2)
\; ,
\label{matrix_element} 
\end{equation}
where $F_{g^* g^* \to \eta'}^2(q_{1t}^2,q_{2t}^2)$
and $F_{\gamma^* \gamma^* \to \eta'}^2(q_{1t}^2,q_{2t}^2)$
are two-gluon and two-photon transition form factors of the $\eta'$
meson, respectively.
It was discussed, e.g. in \cite{KPK2003}, in leading-twist collinear 
approximation. Such an approach is valid for $Q_1^2 = q_{1t}^2 \gg 0$
and $Q_2^2 = q_{2t}^2 \gg 0$. Here we need such a transition form
factor also for $Q_1^2, Q_1^2 \sim 0$.
There is a simple relation between the two-gluon and two-photon
form factors for the quark-antiquark systems 
(see e.g.\cite{BPSS2020,LMS2020,LS2020}).
$\eta'$ meson may have also the two-gluon component in its Fock
decomposition \cite{BM2019}.
The form factor found there can be approximately parametrized as
\begin{equation}
\bar Q^2 F_{g^* g^* \to \eta'}^2(Q_1^2,Q_2^2) \approx 0.2 \pm 0.1 \; \GeV
\; ,
\label{LT_gg_formfactor}
\end{equation}
where $\bar Q^2 = (Q_1^2 + Q_2^2)/2$.
A better approach would be to use their Eqs.(5.13-5.16) with parameters
given there. The result from \cite{KPK2003} is:
\begin{equation}
F(\bar Q^2,\omega) = 4 \pi \alpha_s \frac{f_P}{\bar Q^2}
\frac{\sqrt{n_f}}{N_c} A(\omega) \; .
\label{LT_FF}
\end{equation}
In the factorized (in $\bar Q^2$ and $\omega$) formula:
\begin{equation}
 A(\omega) = A_{q \bar q}(\omega) + \frac{N_c}{2 n_f} A_{gg}(\omega)
\; ,
\label{qqbar_plus_gg}
\end{equation}
where
\begin{eqnarray}
A_{q \bar q}(\omega) &=&  \int_0^1 d x \; \Phi_1(x,\mu_F^2)  
\frac{1}{1 - \omega^2(1-2x)^2} \; ,
\\\
A_{gg}(\omega)       &=&  \int_0^1 d x \; \frac{\Phi_g(x,\mu_F^2)}{x \bar x} 
\frac{1 - 2x}{1 - \omega^2(1-2x)^2} \; 
\end{eqnarray}
and $\Phi_1$ and $\Phi_g$ are singlet and gluon distribution functions,
respectively.
Above 
\begin{equation}
\omega = \frac{Q_1^2 - Q_2^2}{Q_1^2 + Q_2^2} \; .
\label{asymmetry_parameter}
\end{equation}

$\Phi_1$ and $\Phi_g$ undergo QCD evolution \cite{KPK2003} which is
included also in the present paper.

%
%


%
%


\subsection{$F_{\gamma^* \gamma^* \to \eta'}$ form factor}

In Ref.\cite{BGPSS2019} we have shown how to calculate
the transition form factor from the light-cone $Q \bar Q$ wave
function of the $\eta_c$ quarkonium. Here we shall follow the same idea 
but for light quark and light antiquark system.
The flavour wave function of $\eta'$ meson can be approximated as
\cite{DGH1992}
\begin{equation}
| \eta' \rangle \approx \frac{1}{\sqrt{3}}( u \bar u + d \bar d + s \bar s )
\; .
\label{flavour_structure}
\end{equation}
The spatial wave function could be calculated e.g. in potential models.
The momentum wave function can be then obtained as a Fourier transform of
the spatial one. We shall not follow this path in the present study.
Instead we shall take a simple, but reasonable, parametrization
of the respective light-cone wave function.
In principle, each component in (\ref{flavour_structure}) may have 
different spatial as well as momentum wave function.
Here for simplicity we shall assume one effective wave function for 
each flavour component. 
We shall take the simple parametrization of the momentum wave function
\begin{equation}
u(p) \propto \exp \left( p^2/(2 \beta) \right) \; .
\label{momentum_wf}
\end{equation}
The light cone wave function is obtained then via the Terentev's 
transformation (see e.g. \cite{BGPSS2019}).
%
%
We shall use the normalization of the light cone wave function as:
\begin{equation}
\int_0^1 \frac{dz}{z(1-z)} \frac{d^2 k}{16 \pi^3} |\phi(z,k_t)|^2 = 1 \; .
\label{WF_normalization}
\end{equation}
Above
\begin{equation}
\phi(z,k_t) \propto \sqrt{M_{q \bar q}} \exp\left(-p^2/(2 \beta^2)\right)
\end{equation}
and the so-called Terentev's prescription, relating the rest-frame and
light-cone variables, is used:
\begin{equation}
p^2 = \frac{1}{4} \left(M_{q \bar q}^2 - 4 m_{eff}^2  \right) \; .
\end{equation}
Above $M_{q \bar q}$ is the invariant mass of the $q \bar q$ system.

The parameters in the above equations: $m_{eff}$ (hidden in $\phi(z,k_t)$) 
and $\beta$ are in principle free. 
Here we shall take:
\begin{equation}
m_{eff} = (2/3) m_q + (1/3) m_s \; ,
\label{effective_mass}
\end{equation}
where $m_q$ and $m_s$ are constituent masses of light (u,d) and strange
quarks, respectively. Therefore $m_{eff} \sim$ 0.4 GeV.
The weights are from the flavor wave function (\ref{flavour_structure}).
We shall try a few different $\beta$ values in the range (0.4-0.6) GeV.
The normalization constant can be then obtained from the light-cone wave
function normalization.

Having fixed light-cone wave function one can calculate
electromagnetic $\gamma^* \gamma^* \to \eta'$ transition form factor as:
\begin{eqnarray}
F(Q_1^2, Q_2^2) = - \frac{1}{\sqrt{3}} (e_u^2 + e_d^2 + e_s^2) \sqrt{N_c}  \, 4 m_{eff}
&\cdot& \int {dz d^2 \bk \over z(1-z) 16 \pi^3} \psi(z,\bk)  \nonumber \\
\Big\{ 
{1-z \over (\bk - (1-z) \bq_2 )^2  + z (1-z) \bq_1^2 + m_{eff}^2}
&+& {z \over (\bk + z \bq_2 )^2 + z (1-z) \bq_1^2 + m_{eff}^2}
\Big\} \, .
\label{LC_formfactor}
\end{eqnarray}
The $F(0,0)$ is known and can be calculated from the radiative decay
width \cite{BABAR2018}.

The present BABAR data \cite{BABAR2018} are not sufficiently precise 
to get the parameters of our model ($m_{eff}$ and $\beta$).
They could be adjusted in future to precise
experimental data for the $e^+ e^- \to e^+ e^- \eta'$ reaction from
Belle 2

The formula (\ref{LC_formfactor}) can be reduced to a single integral
\begin{eqnarray}
F(Q_1^2, Q_2^2) &=&
 \frac{1}{\sqrt{3}} (e_u^2 + e_d^2 + e_s^2) f_{\eta'} \nonumber \\
&\cdot& \int_0^1 dz 
\Big\{
\frac{(1-z)\phi(z)}{(1-z)^2 Q_1^2 + z(1-z) Q_2^2 + m_{eff}^2}
+
\frac{z \phi(z)}{z^2 Q_1^2 + z(1-z)Q_2^2 + m_{eff}^2}
\Big\}
\label{formfactor_fromDA}
\end{eqnarray}
when introducing so-called distribution amplitudes $\phi(z)$
and so-called decay constant $f_{\eta'}$ (see e.g.\cite{BGPSS2019}).

We shall use also a simple parametrization of the transition form factor
called non-factorized monopole for brevity
\begin{equation}
F^{nf, monopole}(Q_1^2,Q_2^2) = F(0,0)
\frac{\Lambda^2}{\Lambda^2 + Q_1^2 + Q_2^2} \; .
\label{monopole}
\end{equation}
This two-parameter formula can be correctly normalized at $Q_1^2$ = 0
and $Q_2^2$ = 0 \cite{BABAR2018}. It has also correct asymptotic dependence on
$\bar Q^2 = (Q_1^2 + Q_2^2)/2$. This is very similar to the approach
done long ago by Brodsky and Lepage \cite{BL1981} in the case
of neutral pion.

The so-called vector meson dominance model (factorized monopole)
\begin{equation}
F^{VDM}(Q_1^2,Q_2^2) = F(0,0) \frac{m_V^2}{m_V^2+Q_1^2}
\frac{m_V^2}{m_V^2+Q_2^2}
\end{equation}
has incorrectly strong $\bar Q^2$ dependence \cite{BABAR2018}.
We shall compare results obtained with the form factor calculated
with the light-cone wave function (\ref{LC_formfactor}) with the 
parametrization (\ref{monopole}) for $\Lambda$ = 1 GeV.
Results of such a calculation will be treated as a reference ones
for other approaches.

The effect of internal transverse momenta of quarks and antiquarks
in a meson was discussed long ago \cite{Ong1995} postulating some 
wave function of $\pi^0$ in the impact parameter space and 
including suppression due to so-called Sudakov form factor.

In Ref.\cite{KPK2019} the authors tried to adjust the coefficient of the
lowest-order Gegenbauer palynomials to describe the BABAR data 
\cite{BABAR2018} for two virtual photons within the leading-twist 
collinear approximation.
However, the corresponding error bars on expansion coefficients 
are very large.

The two-gluon transition form factor is closely related to
two-photon transition form factor provided the meson is 
of the quark-antiquark type i.e. its wave function as in
Eq.(\ref{flavour_structure}). Then
\begin{equation}
|F_{g^* g^* \to \eta'}(Q_1^2,Q_2^2)|^2 = 
|F_{\gamma^* \gamma^* \to  \eta'}(Q_1^2,Q_2^2)|^2
\frac{g_s^2}{g_{em}^2} \frac{1}{4 N_c (N_c^2 - 1)}
\frac{1}{(<e_q^2>)^2} \; . 
\label{FF_gamgam_to_gg}
\end{equation}
Above $g_{em}^2$ must be taken provided it is included in the definition
of $F_{\gamma^* \gamma^* \to \eta'}$ transition form factor.
Usually it is not.

\begin{figure}
\includegraphics[width=7cm]{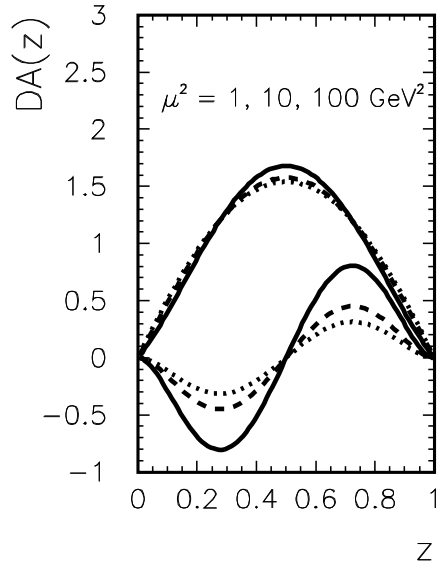}
\caption{$q \bar q$ (upper curves) and $g g$ (lower curves) distribution
  amplitudes for three different
  scales: 1 GeV$^2$ (solid), 10 GeV$^2$ (dashed) and 100 GeV$^2$ (dotted).}
\label{fig:DA_x}
\end{figure}

In Fig.\ref{fig:DA_x} we show $q \bar q$ and $g g$
distribution amplitudes from \cite{KPK2003} for different evolution scales.
Such distribution amplitudes can be used to calculate 
$F_{g^* g^* \to \eta'}$ (see Eq.(\ref{qqbar_plus_gg})) needed
in calculating $\eta'$ production in proton-proton collisions.

\section{Results}

In this section we present our results for $\phi$ and $\eta'$ meson
production.

\subsection{$\phi$ production}

\begin{figure}
\includegraphics[width=8cm]{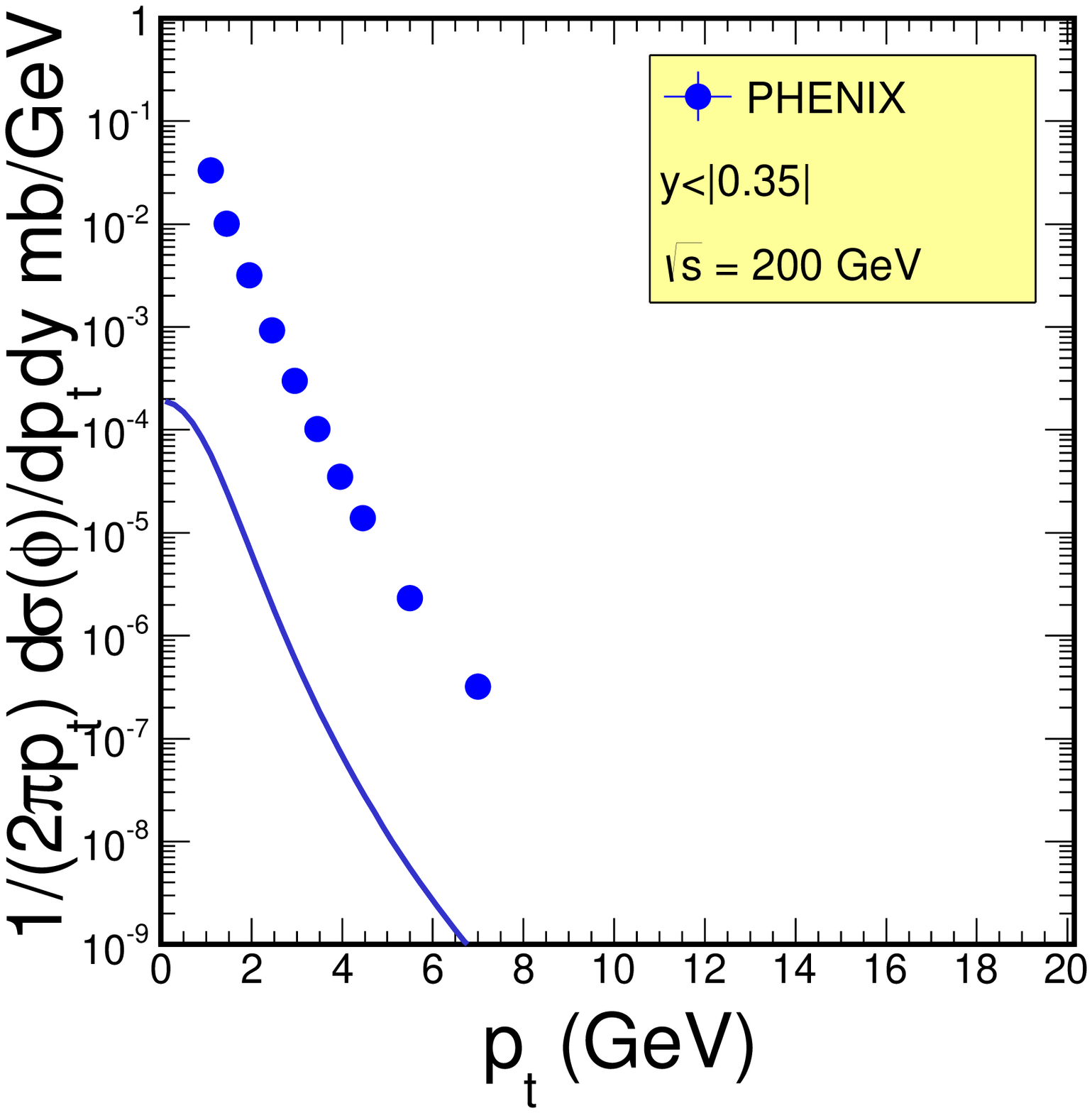}
\includegraphics[width=8cm]{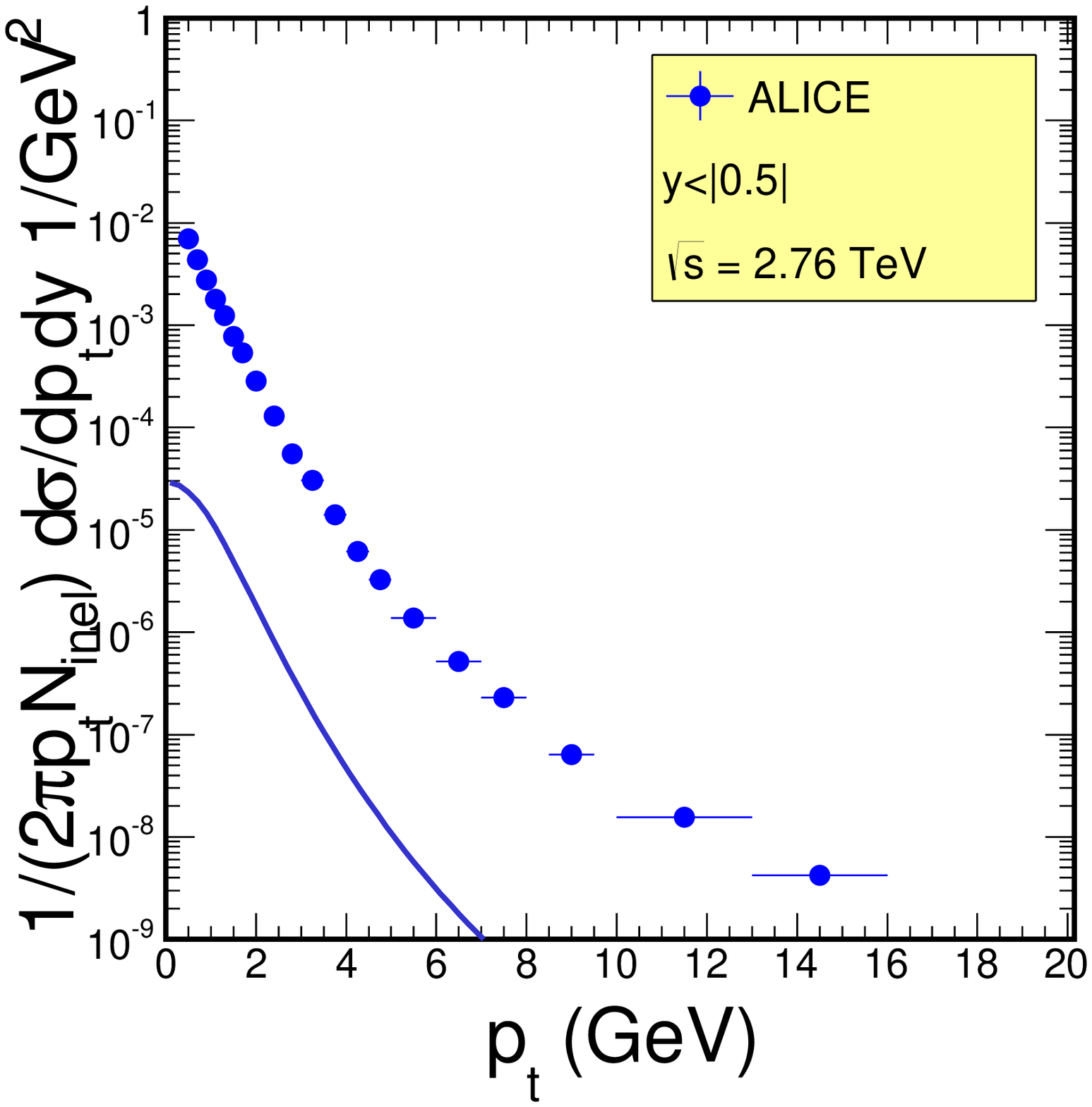} \\
\includegraphics[width=8cm]{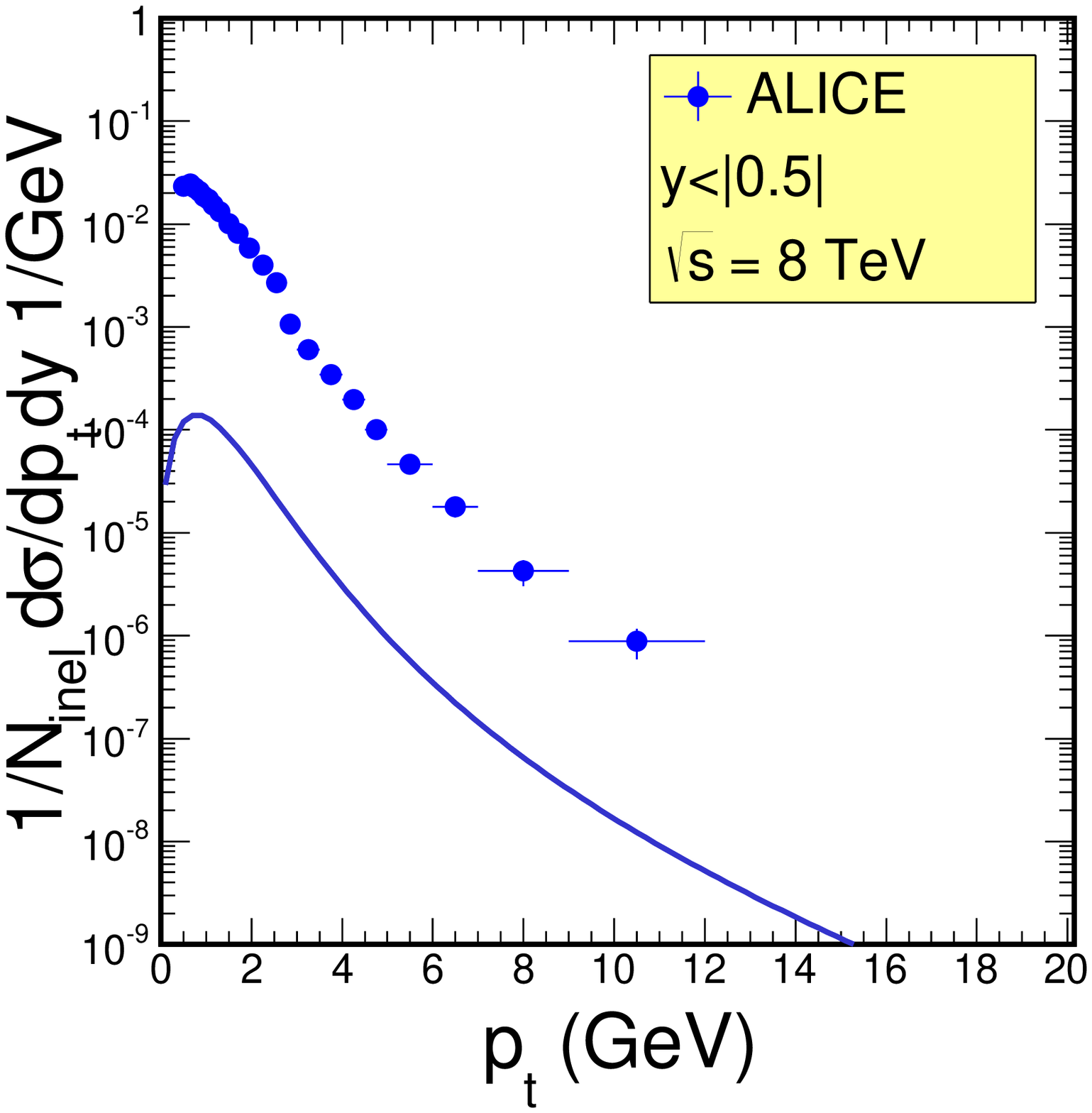}
\caption{Invariant cross section for $\phi$ production
at $\sqrt{s}$ = 200 GeV, 2.76 GeV and 8 TeV. We show the experimental data
of the PHENIX collaboration \cite{PHENIX2011}, ALICE collaboration
\cite{ALICE2017} and results from \cite{kunthia}.
Here $\Psi(0)$ was calculated from Eq.(\ref{Psi(0)}). 
}
\label{fig:phi_invariant_cs}
\end{figure}

In this subsection we show the cross section for $\phi$ meson
production for $\sqrt{s}$ = 200 GeV, $\sqrt{s}$ = 2.76 GeV and 
$\sqrt{s}$ = 8 TeV (see Fig.\ref{fig:phi_invariant_cs}).
Our results are shown together with the PHENIX \cite{PHENIX2011} and ALICE
\cite{ALICE2017,kunthia} experimental data, respectively.\\
For each considered case the result of calculation is below the
experimental data. This suggests that the gluon-gluon fusion is not the
dominant production mechanism of $\phi$ meson production.
The fragmentation mechanism was considered in \cite{SIM2014,SI2020}
and it may be the dominant mechanism of $\phi$ meson production.

\subsection{$F_{\gamma^* \gamma^* \to \eta'}$ form factor}

Before presenting our results for the $\eta'$ production we wish to show
our results for the $F_{\gamma^* \gamma^* \to \eta'}$ form factor.

\begin{figure}
\includegraphics[width=7cm]{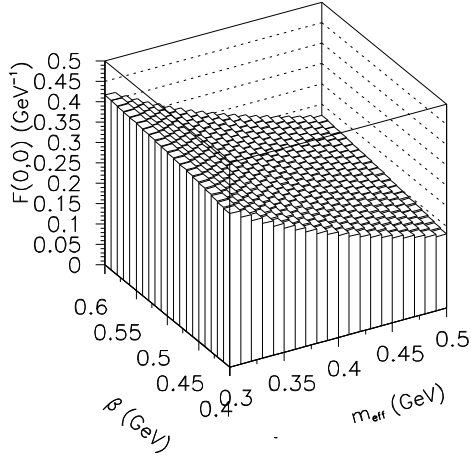}
\caption{$F_{\gamma^* \gamma^* \to \eta'}(0,0)$ as a function of $\beta$
and $m_{eff}$.}
\label{fig_F00}
\end{figure}

We will start with our results obtained from the LCWF for 
$F_{\gamma \gamma \to \eta'}(0,0)$.
In Fig.\ref{fig_F00} we present $F_{\gamma \gamma \to \eta'}(0,0)$
as a function of $\beta$ and $m_{eff}$. There is a small dependence
on the parameters. The experimental value is
\begin{equation}
F_{\gamma^* \gamma^* \to \eta'}(0,0) = 
\sqrt{\frac{4 \Gamma_{\eta' \to 2 \gamma}}{\pi \alpha_{em}^2 m_{\eta'}^3}}
= 0.342 \pm 0.006 \; \GeV^{-1} \; .
\end{equation}
A broad range of $\beta$ and $m_{eff}$ is allowed taken into account
the simplicity of our approach.

\begin{figure}
\includegraphics[width=7cm]{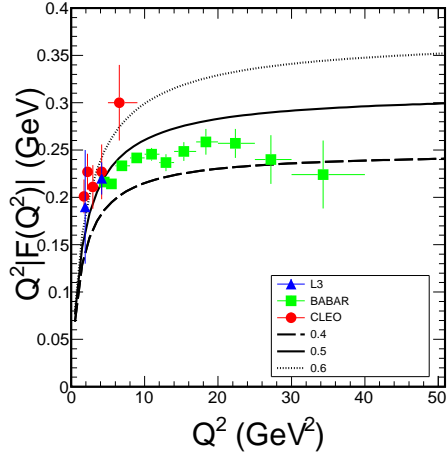}
\caption{$Q^2 F(Q^2)$ as a function of one photon virtuality.
We show results for $m_{eff}$ = 0.4 GeV and for different
values of $\beta$ = 0.4, 0.5, 0.6 GeV (from bottom to top).
For comparison we show experimental data from \cite{CLEO,L3,BABAR2011}.}
\label{fig:Q2F}
\end{figure}

Most of the studies on transition form factors has been concentrated on
the case of only one virtual photon. In Fig.\ref{fig:Q2F}
we show $Q^2 F_{\gamma^* \gamma^* \to \eta'}(Q^2)$ for different 
values of model parameters $\beta$ = 0.4, 0.5, 0.6 GeV.
In leading twist approach, without QCD evolution of distribution
amplitudes, one should get a constant at large photon virtualities.
In our approach this happens at extremely large virtualities.
Below $Q^2 <$ 50 GeV our model clearly contains higher twists.
In collinear leading twist approach the rise of 
$Q^2 F_{\gamma^* \gamma^* \to \eta'}(Q^2)$ is 
caused by the evolution (see e.g. \cite{KPK2013}).

\begin{figure}
\includegraphics[width=7cm]{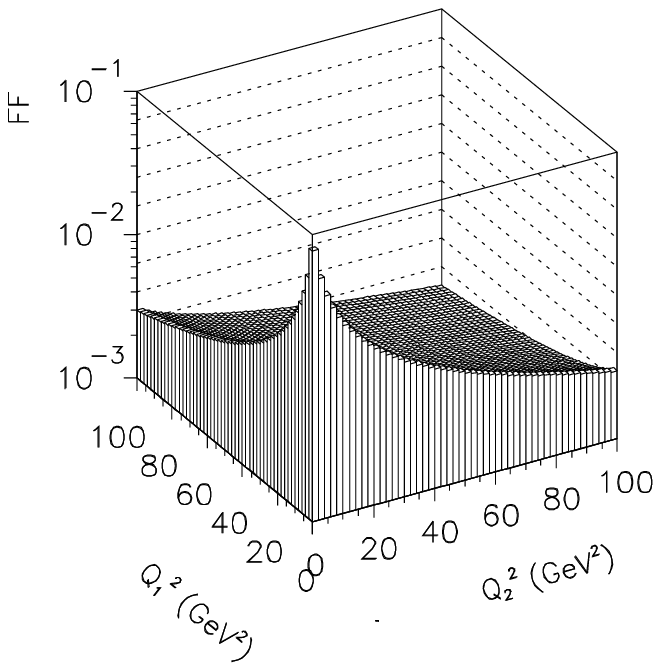}
\includegraphics[width=7cm]{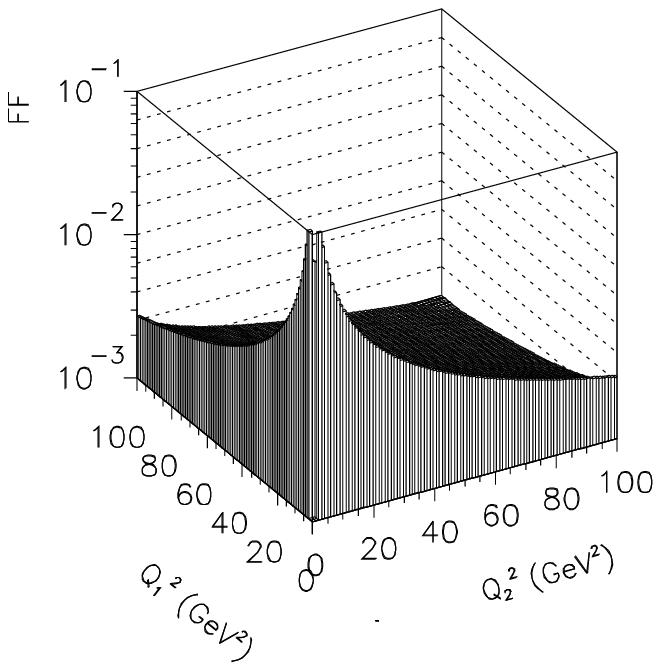}
\caption{$F_{\gamma^* \gamma^* \to \eta'}(Q_1^2,Q_2^2)$ obtained with 
the light-cone wave function
for $m_{eff}$ = 0.4 GeV and $\beta$ = 0.5 GeV (left panel)
and the leading-twist result from Ref.\cite{KPK2019} (right panel).}
\label{fig:F_Q12Q22}
\end{figure}

In Fig.\ref{fig:F_Q12Q22} we show the two-photon $\eta'$ form factor
as a function of both photon virtualities.
The form factor drops quickly from 
$F_{\gamma^* \gamma^* \to \eta'}(0,0)$ in the region 
$Q_1^2 <$ 10 GeV$^2$, $Q_2^2 <$ 10 GeV$^2$.
The change beyond this region is rather mild.
We show our result obtained using the light-cone wave function (left
panel) and for comparison the result from Ref.\cite{KPK2019} (right panel).
The leading-twist result is realiable only for larger virtualities.
Both the results are similar for larger virtualities.

\begin{figure}
\includegraphics[width=8cm]{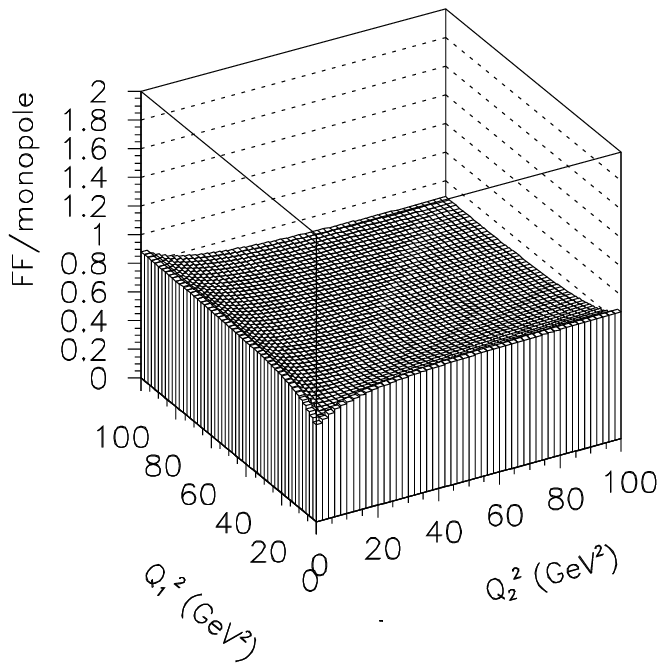}
\includegraphics[width=8cm]{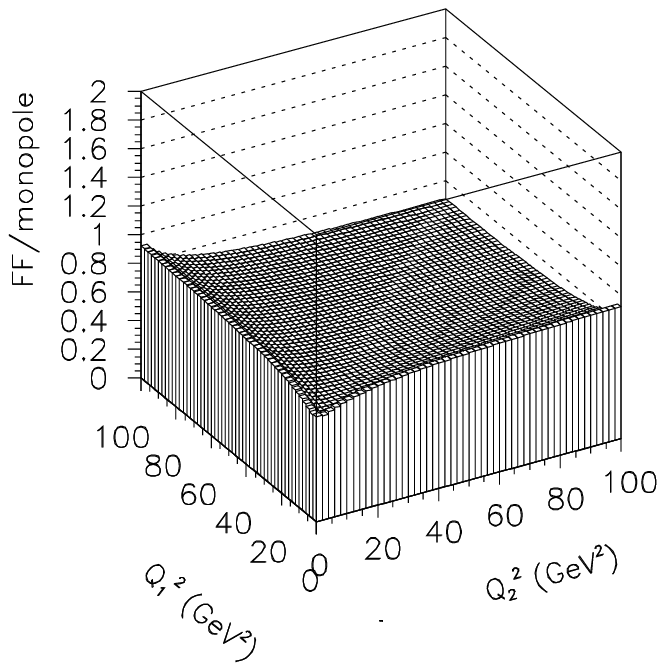}
\caption{$R(Q_1^2,Q_2^2)$ (see Eq.(\ref{ratio})) with the 
$\gamma^* \gamma^* \to \eta'$ form factor
calculated with $\eta'$ light-cone wave function (left panel).
In this calculation $m_{eff}$ = 0.4 GeV and $\beta$ = 0.5 GeV
is used for example.
In the right panel we show similar ratio obtained from
Eq.(\ref{formfactor_fromDA}) with asymptotic distribution amplitude. 
}
\label{fig:R_Q12Q22}
\end{figure}

In order to better visualize our result we show in 
Fig.\ref{fig:R_Q12Q22} also the ratio:
\begin{equation}
R(Q_1^2,Q_2^2) = F^{LC}(Q_1^2,Q_2^2) / F^{nf, monopole}(Q_1^2,Q_2^2)
\; 
\label{ratio}
\end{equation}
and similar obtained when using formula (\ref{formfactor_fromDA}).

We observe that the form factor calculated from (\ref{LC_formfactor}) 
deviates only slightly from the simple parametrization (\ref{monopole}).
The two parametrizations almost coincide in the broad range of 
$(Q_1^2,Q_2^2)$.
A similar result is obtained when using Eq.(\ref{formfactor_fromDA})
with asymptotic distribution amplitude.

\begin{figure}
\includegraphics[width=8cm]{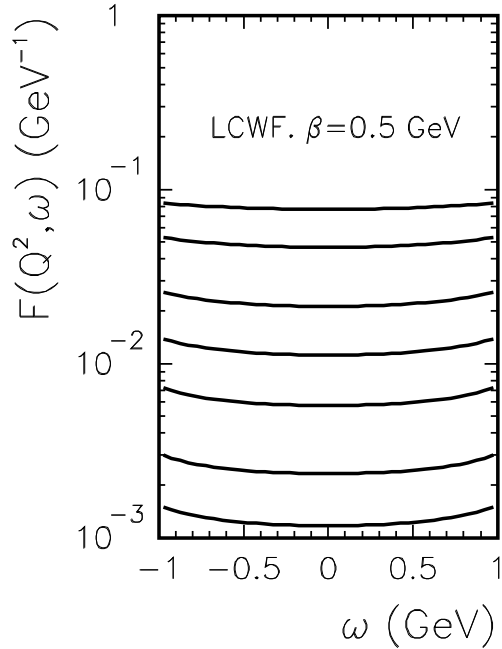}
\caption{The $\gamma^* \gamma^* \to \eta'$ form factor
as a function of $\omega$ for a fixed values of $\bar Q^2$ 
(from top to bottom: 2, 5, 10, 20, 50, 100 GeV$^2$)
calculated with $\eta'$ light-cone wave function.
In this calculation $m_{eff}$ = 0.4 GeV and $\beta$ = 0.5 GeV. 
}
\label{fig:F_omega}
\end{figure}

In Fig.\ref{fig:F_omega} we show the two-photon transition form factor
 as a function of asymmetry parameter $\omega$ 
(see Eq.(\ref{asymmetry_parameter}))
for different values of $\bar Q^2$ specified in the figure caption.
In contrast to the non-factorizable monopole form factor
(\ref{monopole}) we get some dependence on asymmetry parameter
$\omega$. This dependence is somewhat similar to the result of 
Ref.\cite{KPK2019}.
In contrast to \cite{KPK2019} our dependence on $\omega$ is not
universal, i.e. different for different values of $\bar Q^2$.
 
The two-photon form factor will be transformed to two-gluon form factor
and the latter will be used in the calculation of $\eta'$ production.
For this purpose first a grid for $F_{\gamma^* \gamma^* \to \eta'}$ 
in the $(\bar Q^2,\omega)$ plane is prepared.
The grid is then used in the interpolation when calculating
differential distributions of $\eta'$ meson
in proton-proton collisions.

\subsection{$\eta'$ production}

In this subsection we discuss the $\eta'$ production considering the
 simple gluon-gluon fusion mechanism illustrated in Fig.\ref{fig:gg_etap}.



%
%

\begin{figure}
\includegraphics[width=8cm]{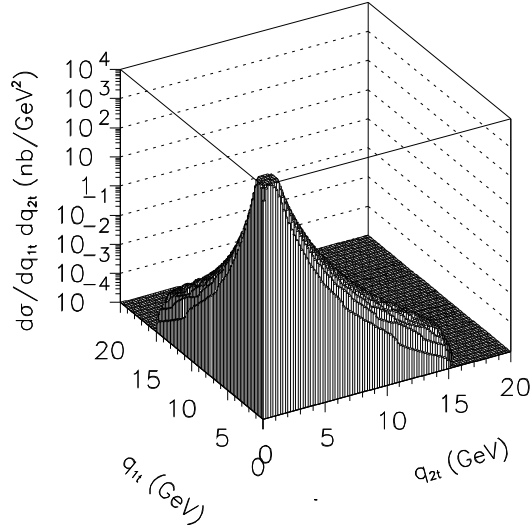}
\caption{
Two-dimensional map in ($q_{1t},q_{2t}$) for the full range
of $\eta'$ transverse momentum. 
Here $\sqrt{s}$ = 200 GeV and the KMR UGDF was used.
}
\label{fig:dsig_dk1tdk2t_pt_full}
\end{figure}

What are typical gluon transverse momenta for hadronic 
$g^* g^* \to \eta'$ process.
In Fig.\ref{fig:dsig_dk1tdk2t_pt_full} we show the 
distribution of the cross section
integrated over $\eta'$ transverse momenta in the $(q_{1t},q_{2t})$.
Both small and large gluon virtualities enter into the $p_t$-integrated
cross section.

\begin{figure}
\includegraphics[width=7cm]{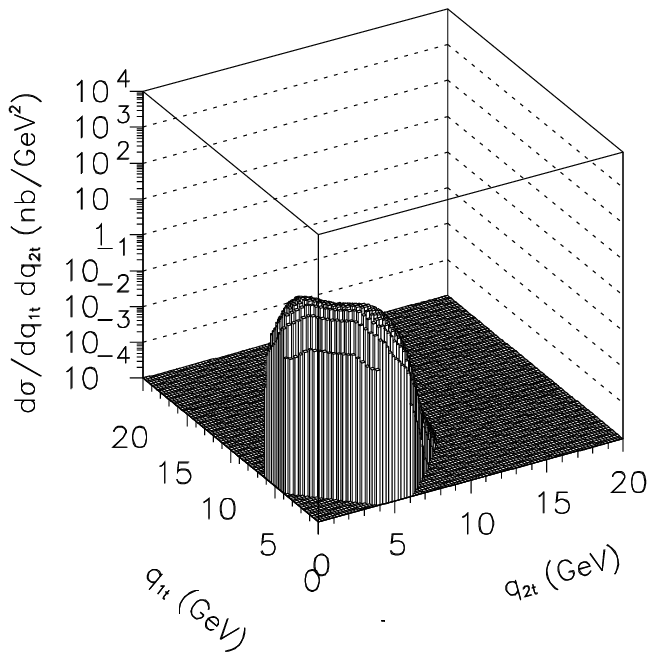}
\includegraphics[width=7cm]{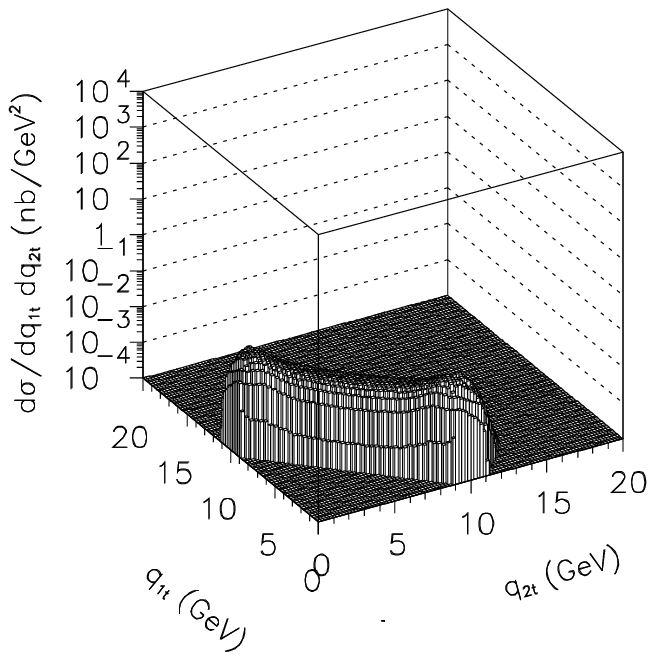}
\caption{
Two-dimensional map in ($q_{1t},q_{2t}$) for two different ranges
of $\eta'$ transverse momentum: 4 GeV $< p_t <$ 6 GeV (left panel) 
and 9 GeV $< p_t <$ 11 GeV (right panel). 
Here $\sqrt{s}$ = 200 GeV and the KMR UGDF was used.
}
\label{fig:dsig_dk1tdk2t_pt_regions}
\end{figure}

In Fig.\ref{fig:dsig_dk1tdk2t_pt_regions} we show similar distributions
as above but for two different regions of meson transverse momentum $p_t$. 
The larger $p_t$ the larger gluon transverse momenta enter into the game.

\begin{figure}
\includegraphics[width=6cm]{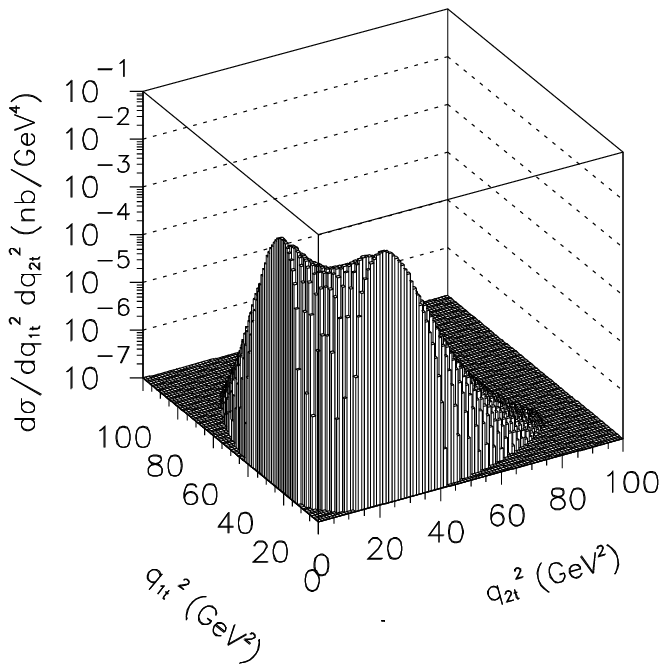}
\includegraphics[width=6cm]{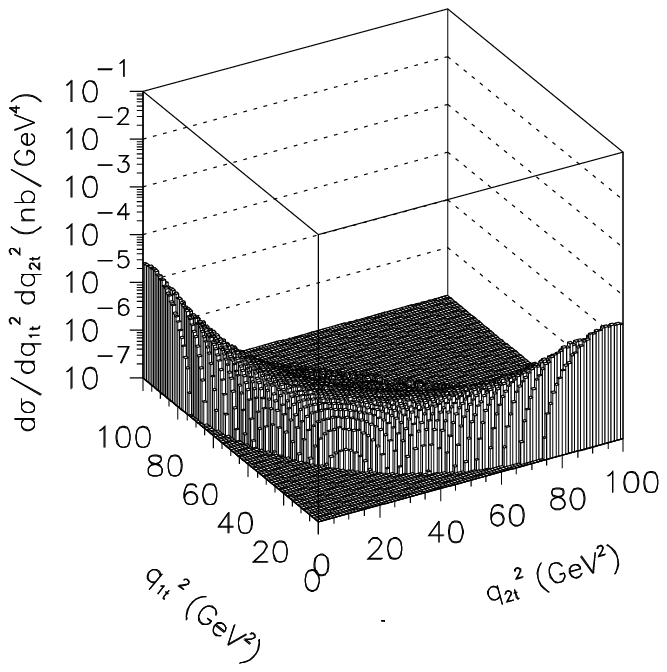}
\caption{
Two-dimensional map in ($q_{1t}^2,q_{2t}^2$) for 
4.5 GeV $< p_t <$ 5.5 GeV (left panel) and 
9.5 GeV $< p_t <$ 10.5 GeV (right panel). 
Here $\sqrt{s}$ = 200 GeV. In this caluculation the KMR UGDF was used
and the light-cone wave function with $\beta$ = 0.5 GeV.
}
\label{fig:dsig_dk1t2dk2t2_pt_fixed}
\end{figure}

Fig.\ref{fig:dsig_dk1t2dk2t2_pt_fixed} shows a similar distributions 
but in the $(q_{1t}^2,q_{2t}^2)$ space used usually for presentation
of transition form factors.
One can observe that at large transverse momentum $p_t$ one is sensitive
to the region of $Q_1^2$ very small and $Q_2^2$ very large or
   $Q_1^2$ very large and $Q_2^2$ very small.
These are regions relevant for the leading-twist collinear approach
to two-gluon transition form factor. This is also the region 
of the phase space where
the meson light-cone approach gives a small relative enhancement 
compared to the naive monopole parametrizations (see Fig.\ref{fig:R_Q12Q22}).
We observe that at $p_t \sim$ 10 GeV the gluon virtualities
$Q_1^2 >$ 20 GeV$^2$ or $Q_2^2 >$ 20 GeV$^2$ are clearly in the domain 
of leading-twist approach \cite{KPK2003}.

\begin{figure}
\includegraphics[width=7cm]{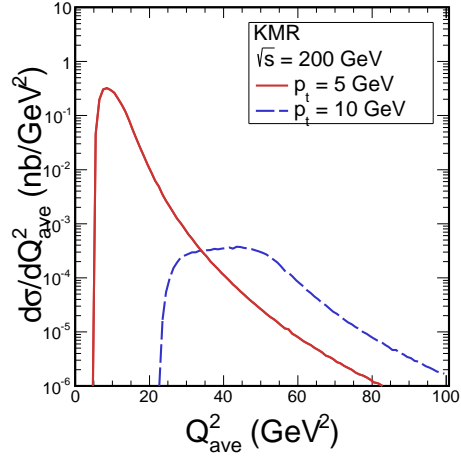}
\caption{
Distribution in $Q_{ave}^2 = \bar Q^2$ for the two distinct cases 
from the previous
figure: $p_t$ = 5 $\pm$ 0.5 GeV (solid line) and
        $p_t$ = 10 $\pm$ 0.5 GeV (dashed line).
Here $\sqrt{s}$ = 200 GeV. In this caluculation the KMR UGDF was used
and the light-cone wave function with $\beta$ = 0.5 GeV.
}
\label{fig:dsig_dQ2ave}
\end{figure}

This is shown better in Fig.\ref{fig:dsig_dQ2ave} where
we display distribution in $\bar Q^2$. Only large $\bar Q^2$ occur
for $p_t >$ 10 GeV. This situation is generic, independent
of the form factor used.

\begin{figure}
\includegraphics[width=8cm]{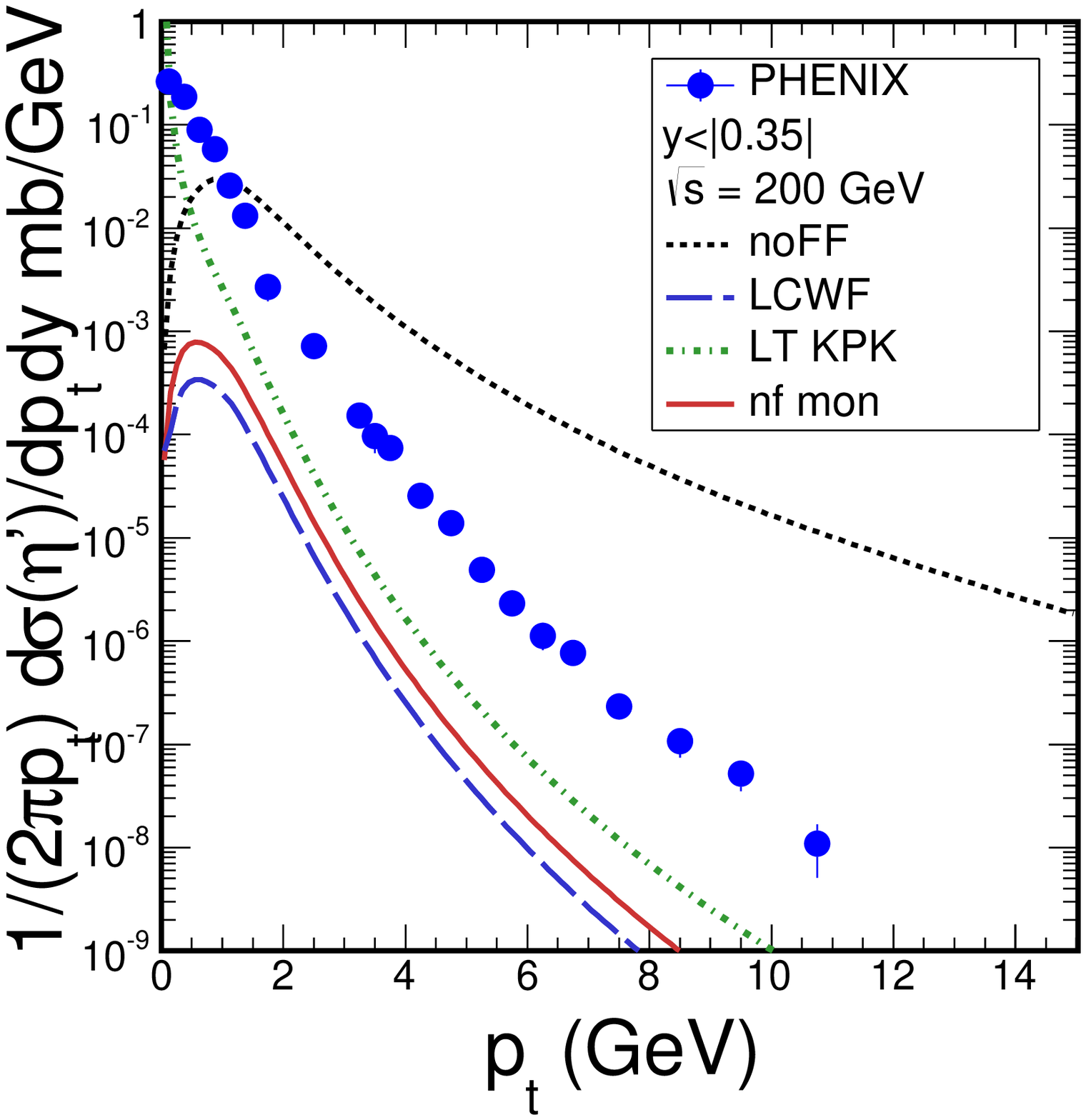}
\includegraphics[width=8cm]{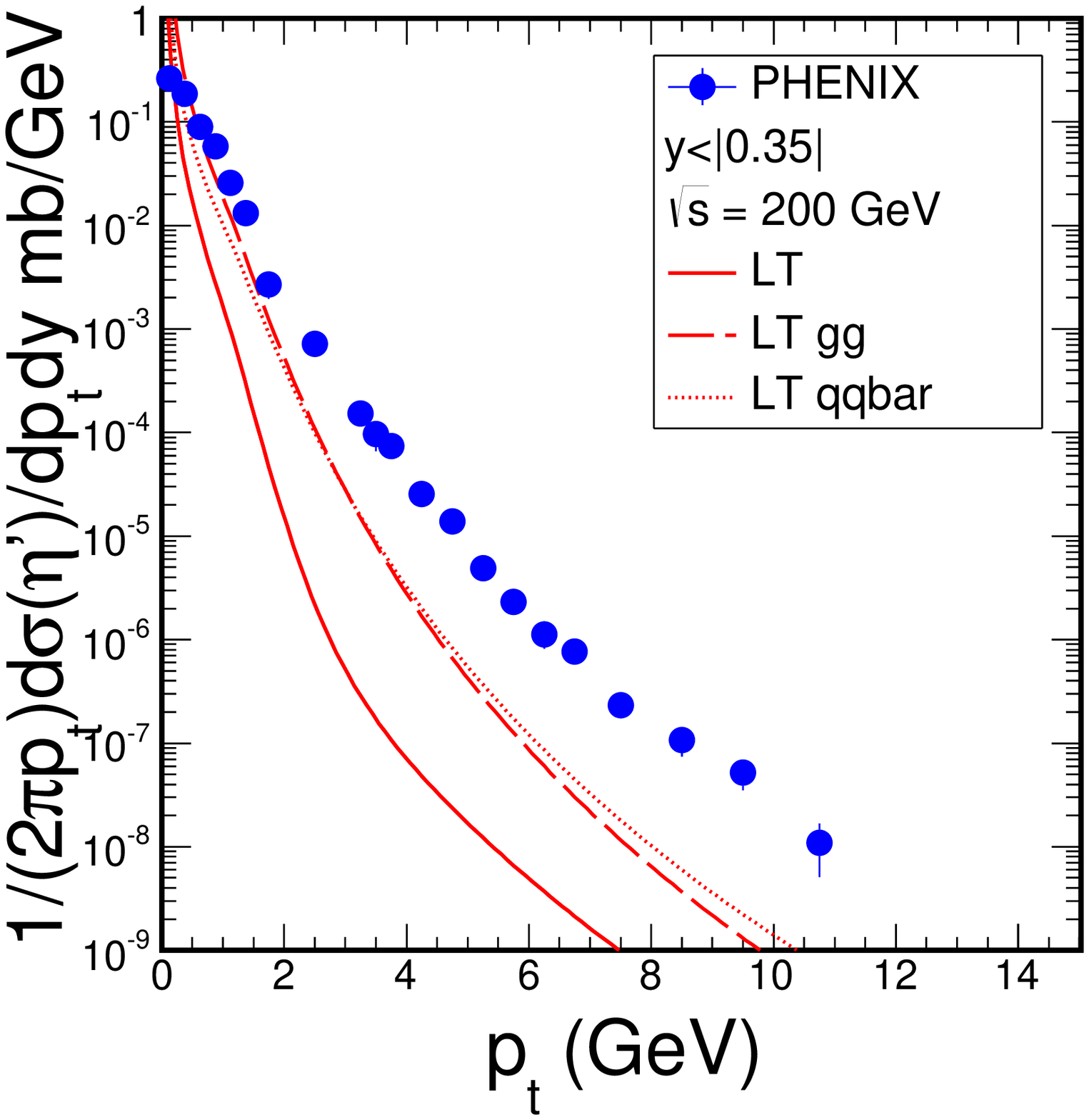}
\caption{Invariant cross section as a function of meson transverse momentum. 
Here $\sqrt{s}$ = 200 GeV and the KMR UGDF was used in the calculation.
In the left panel results for the nonfactorized monopole, LCWF with
$\beta$ = 0.5 GeV, and simple LT parametrization.
In the right panel we show results obtained using distribution
amplitudes from \cite{KPK2003}. We show the full result as well as
result when only $q \bar q$ or only $g g$ components in 
(\ref{qqbar_plus_gg}) are included.
} 
\label{fig:dsig_dyd2pt_phenix}
\end{figure}

Finally in Fig.\ref{fig:dsig_dyd2pt_phenix} we show invariant cross
section for the $\eta'$ meson production for the RHIC energy
$\sqrt{s}$ = 200 GeV relevant for the PHENIX experiment
\cite{PHENIX2011}.
In the left panel we show different results:\\
(a) with non factorized monopole form factor (solid line),\\
(b) with the form factor calculated from the LCWF with 
$\beta$ = 0.5 GeV (dashed line),\\
(c) leading twist parametrization (\ref{LT_gg_formfactor}) of 
the $F_{g^* g^* \to \eta'}$ result from \cite{KPK2003}
(dash-dotted line),\\
(d) without $F_{g^* g^* \to \eta'}$, except of normalization constant
(dotted line).\\
The result obtained with the leading-twist parametrization 
(\ref{LT_gg_formfactor}) can be taken serious only for 
$p_t >$ 5 GeV, when $Q_1^2$ or $Q_2^2$ are bigger than 5 GeV$^2$.
We also used the formalism of collinear distribution amplitudes from
\cite{KPK2003}, including their QCD evolution, to calculate
$F_{g^* g^* \to \eta'}(q_{1t}^2,q_{2t}^2)$ from evolved
$q \bar q$ and $g g$ distribution amplitudes.
The evolution scale of distribution amplitudes is taken as
$\mu^2 = \bar Q^2 + \mu_0^2$, where $\mu_0^2$ = 1 GeV$^2$ is used
in our calculation.
The corresponding result is shown by the red thick solid line.
The line is below other lines in the region where the experimental
data exist. For comparison we show somewhat arbitrarily also result with
$q \bar q$ alone (red thick dashed line) and $g g$ alone (red thick
dotted line) in Eq.(\ref{qqbar_plus_gg}). Both the results are much 
bigger than the result when both components are included coherently. 
Clearly a strong destructive interference effect of both contributions 
is observed. The opposite 
sign of the $g g$ distribution amplitude would cause constructive 
interference in Eq.(\ref{qqbar_plus_gg}).

In all cases we get less cross section than measured by the PHENIX
collaboration at RHIC. This suggests that the gluon-gluon fusion
is probably not the dominant mechanism of $\eta'$ production,
at least in the measured region of transverse momenta.
A natural candidate is fragmentation process, which was
not discussed in the literature so far in the context of $\eta'$ production.

Neglecting the form factor at all leads to overestimation of the cross
section at large transverse momenta of $\eta'$ (see the dotted line
in the left panel of Fig.\ref{fig:dsig_dyd2pt_phenix}).
Such a result one could expect in the TMD (transverse momentum dependent
gluon distributions) approach \cite{Echevarria:2019ynx} where 
the incoming gluons should be taken on mass shell. The present result
shows therefore shortcomings of the TMD approach in the context 
of meson production.

\begin{figure}
\includegraphics[width=8cm]{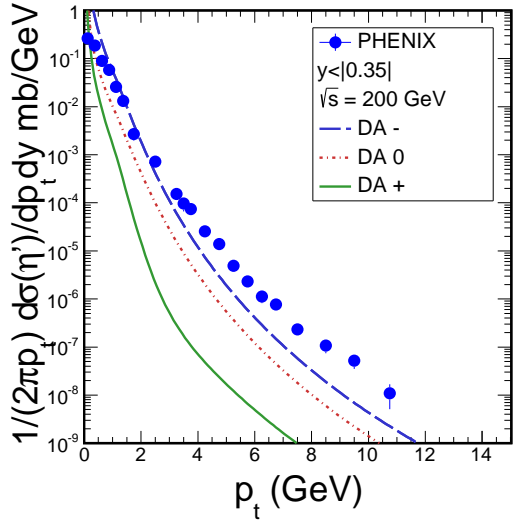}
\caption{Invariant cross section as a function of meson transverse
  momentum in the approach with distribution amplitudes and
different initial $\Phi_{gg}$. 
Here $\sqrt{s}$ = 200 GeV and the KMR UGDF was used in the calculation.
}
\label{fig:different_initial_gg}
\end{figure}

To better illustrate the role of the initial $g g$ component 
in the approach with distribution amplitudes
in Fig.\ref{fig:different_initial_gg} we show the final 
(including QCD evolution) result with different initial $g g$ component:
as in \cite{KPK2003} (plus), with opposite sign (minus) and
with initial $g g$ component put to zero. The final results are
quite different. We conclude that the $\eta'$ transverse momentum distribution
is very sensitive to the unknown nonperturbative $g g$ distribution
amplitude.

\begin{figure}
\includegraphics[width=8cm]{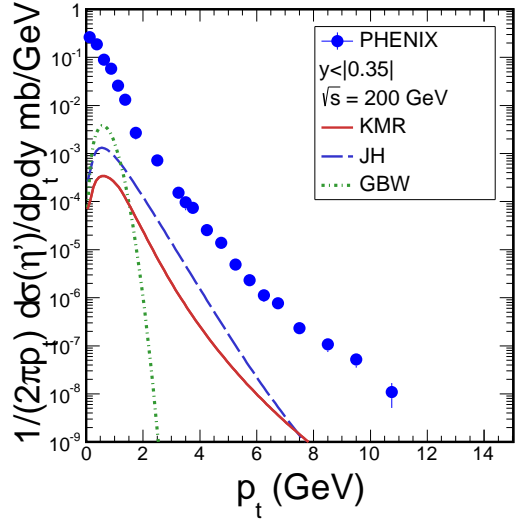}
\caption{Invariant cross section as a function of meson transverse momentum. 
Here $\sqrt{s}$ = 200 GeV.
We show results with the KMR (solid line), Jung-Hautmann (dashed line)
and GBW (dash-dotted line) UGDFs.
In this calculation the form factor based on the LCWF with
$\beta$ = 0.5 GeV is used for illustration.
}
\label{fig:dsig_dyd2pt_phenix_ugdf}
\end{figure}

So far we used only one unintegrated gluon distribution.
In Fig.\ref{fig:dsig_dyd2pt_phenix_ugdf} we compare results obtained
using different UGDFs. The result obtained with the Jung-Hautmann UGDF
is similar to that obtained with the KMR UGDF. 
The GBW UGDF gives sizeable cross section only at low $\eta'$ transverse
momenta as it does not include higher order perturbative effects.

\begin{figure}
\includegraphics[width=8cm]{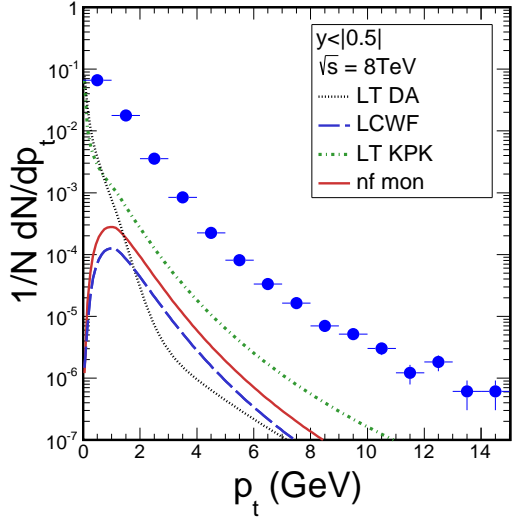}
\caption{Number of $\eta'$ mesons per event as a function of meson 
transverse momentum. 
Here $\sqrt{s}$ = 8 TeV and the KMR UGDF was used in the calculation.
The result of the Lund string model simulations is shown
as ``data points'' for comparison.} 
\label{fig:dN_dpt_ALICE}
\end{figure}

What about larger energies ?
The number of $\eta'$ per event as a function of $\eta'$ transverse
momentum is shown in Fig.\ref{fig:dN_dpt_ALICE} for $\sqrt{s}$ = 8 TeV.
We show the result for the non-factorized monopole (\ref{monopole}) 
two-photon transition form factor (solid line), light-cone wave function
with $\beta$ = 0.5 GeV (dashed line), the result with 
the simple leading twist parametrization (\ref{LT_gg_formfactor}) of 
the two-gluon transition form factor and the results of collinear
approach with evolution of distribution amplitudes (see Eq.(\ref{LT_FF})).
For comparison we show also result from the Lund string model.
The two-gluon mechanism gives much smaller cross section 
than that from the Lund-string model. So even at the LHC we do not 
find any region of the phase space ($y, p_t$) where the two-gluon fusion
is the dominant mechanism of $\eta'$ production.

\section{Conclusions}

In this study we have considered production of two ($\phi$ and $\eta'$)
isoscalar mesons with hidden strangeness via gluon-gluon fusion 
in proton-proton collisions for different collision energies relevant 
for RHIC and the LHCb. The calculations have been performed within
$k_t$-factorization approach with the KMR UGDF which is known
to include effectively higher-order corrections \cite{MS2016,MS2019}.

For the $\phi$ production we extend the calculation performed earlier 
for the $J/\psi g$ production by using effective spatial wave function
at the origin 
$R_{s \bar s}(0)$ which can be estimated from the decay 
$\phi \to e^+ e^-$ by adjusting it to experimental branching ratio.
Having found the parameter we have compared results of our calculation
with the PHENIX and ALICE experimental data. In both cases
the calculated cross section stays below the experimental data
by two (PHENIX) and by one (ALICE) order of magnitude.
This shows that another mechanism is more important.
The fragmentation of $s / {\bar s} \to \phi$ is a natural candidate.

Inspired by the successful description
of $\eta_c$ production in proton-proton collisions \cite{BPSS2020}, 
here we have considered the $g^* g^* \to \eta'$ fusion with 
off-shell initial gluons.
The coupling can be described by the two-gluon nonperturbative
transition form factor. For the quark-antiquark states the latter object
is closely related to the two-photon transition form factor,
studied theoretically and measured by the CLEO, L3 and BABAR collaborations.

The two-photon form factor has been calculated using a light-cone 
wave function for different values of model parameters.
The so-obtained form factor has been compared with a simple
non-factorized monopole parametrization as well as the results 
obtained recently by Kroll and Passek-Kumericki in the leading-twist 
collinear NLO approach.

The two-photon form factors have been translated to the two-gluon ones
assuming the dominance of the quark-antiquark components in the
Fock $\eta'$ wave function expansion. Then it was used in the
$k_t$-factorization approach to calculate the cross section for $\eta'$
production in $p p$ collisions.
The results have been compared with the PHENIX experimental data. 
In spite of the expectation of the community
the calculated cross section is definitely smaller than the measured one
obtained by the PHENIX collaboration.
The situation may improve at larger energies but the relevant cross
section at the LHC was not measured so far.
We have presented our predictions for the LHC and has compared
our two-gluon fusion result with the result form the Pythia generator.
For $\sqrt{s}$ = 8 TeV the cross section from the Lund string model is
much above that for the two-gluon fusion mechanism.
Respective data from the ALICE collaboration would be very important 
to clarify the situation.

\vskip+5mm
{\bf Acknowledgments}\\

A.S. is indebted to Wolfgang Sch\"afer for long-standing collaboration
on quarkonia production.
We are also indebted to Francois Fillion-Gourdeau for pointing to us
Ref.\cite{FJ2009}, Kornelija Passek-Kumericki for providing us
the two-photon $\eta'$ form factor from their leading-twist analysis
in Ref.\cite{KPK2019} and interesting discussion on
transition form factors, Arvind Khuntia for providing us
experimental data of the ALICE collaboration for $\phi$ production 
at $\sqrt{s}$ = 8 TeV presented in his PhD thesis, Jacek Biernat and 
Jacek Otwinowski for providing us results of the Lund-string model 
generator Phytia, Francesco Giacosa for a discussion on gluonic 
components in mesons, and Jacek Oko{\l}owicz for carefull reading
of this manuscript.
This study was partially supported by the Polish National Science Center
grant UMO-2018/31/B/ST2/03537 and by the Center for Innovation and
Transfer of Natural Sciences and Engineering Knowledge in Rzesz{\'o}w.



\end{document}